\documentclass{ws-procs9x6}

\begin{document}

\title{Spin Sum Rules and Polarizabilities: \\
Results from Jefferson Lab}

\author{Jian-ping Chen}

\address{Thomas Jefferson National Accelerator Facility \\ 
12000 Jefferson Ave., Newport News, Virginia 23606, USA \\
$^*$E-mail: jpchen@jlab.org}

\begin{abstract}
The nucleon spin structure has been an active, exciting and intriguing subject
of interest for the last three decades.
Recent experimental data on nucleon spin structure at low 
to intermediate momentum transfers 
provide new information in
the {\it confinement} regime and the
 transition region from the {\it confinement} regime to the {\it
asymptotic freedom} regime. 
New insight is gained by
exploring moments of spin structure functions and their corresponding sum rules (i.e.
the generalized Gerasimov-Drell-Hearn, Burkhardt-Cottingham and Bjorken). The Burkhardt-Cottingham sum rule is verified to 
good accuracy.  
The spin structure moments data are compared with Chiral Perturbation Theory calculations at low momentum transfers. It is found that chiral perturbation calculations agree reasonably well 
with the first moment of the spin structure function $g_1$   
at momentum transfer of 0.05 to 0.1 GeV$^2$ but fail to reproduce the neutron data in the case 
of the generalized polarizability $\delta_{LT}$ (the $\delta_{LT}$ puzzle). 
New data have been taken on
the neutron ($^3$He), the proton and the deuteron  at very low $Q^2$ down to 
0.02 GeV$^2$. They will provide benchmark tests of Chiral dynamics in the 
kinematic region where the Chiral Perturbation theory is expected to work.  

\end{abstract}


\bodymatter

\section{Introduction}	
In the last twenty-five years the study of the spin structure of the nucleon led to a very productive
experimental and theoretical activity with exciting results and new challenges\cite{Fji}. This
investigation has included a variety of aspects, such as
testing QCD in its perturbative regime 
{\emph {via}} spin sum rules (like the Bjorken sum rule\cite{bjo66}) and understanding how the spin of the nucleon is
built from the intrinsic degrees of freedom of the theory, quarks and gluons. 
Recently, results from a new generation of experiments performed at
Jefferson Lab seeking to probe the theory in its non-perturbative 
and transition regimes
 have reached a mature state.  The low momentum-transfer results offer insight in 
a region known for the collective behavior of the nucleon constituents and their interactions.
Furthermore, 
distinct features seen in the nucleon response to the electromagnetic probe, depending on the 
resolution  of the probe, point clearly to different 
regimes of description, i.e. a scaling regime where quark-gluon correlations are suppressed 
versus a coherent regime where long-range interactions give rise to the 
static properties of the nucleon.

In this talk we describe an investigation\cite{e94010,eg1p,eg1d,BJ,eg1b,chen}
 of the spin structure of the nucleon through  
the measurement of the helicity-dependent photoabsorption cross sections or asymmetries using virtual photons  
across a wide resolution spectrum. These observables are used to extract the
spin structure 
functions $g_1$ and $g_2$ and evaluate their moments. These moments are powerful tools 
to test QCD sum rules and Chiral Perturbation Theory calculations.

\section{Sum rules and Moments}

Sum rules involving the spin structure of the nucleon offer an important opportunity to study QCD. In recent years
the Bjorken sum rule at large $Q^2$ (4-momentum transfer squared) and 
the Gerasimov, Drell and Hearn (GDH) sum rule\cite{GDH} at $Q^2=0$
have attracted large experimental and theoretical efforts\cite{dre01} that have provided us with rich information. 
Another type of sum rules, such as the generalized GDH sum rule\cite{ji01} or
the polarizability sum rules\cite{dre}, relate
the moments of the spin structure functions to real or virtual 
Compton amplitudes, which can be calculated theoretically. 
These sum rules are based on ``unsubtracted'' dispersion relations
and the optical theorem. 

Considering the forward spin-flip doubly-virtual Compton scattering (VVCS) amplitude $g_{TT}$ and assuming it has an appropriate 
convergence behavior at high energy,  
an unsubtracted dispersion relation leads to the following equation 
for\cite{dre,chen}
$g_{TT}$:
\begin{equation}
{\rm Re}[{g}_{TT}(\nu,Q^2)-g^{pole}_{TT}(\nu,Q^2)]
=
(\frac{\nu}{2 \pi^2}){\cal P}\int^{\infty}_{\nu_0}\frac{K(\nu',Q^2)
\sigma_{TT}(\nu',Q^2)}{\nu'^2-\nu^2}d\nu', 
\end{equation}
where $g_{TT}^{pole}$ is the nucleon pole (elastic) contribution, ${\cal P}$ 
denotes the principal value integral and
$K$ is the virtual photon flux factor.
The lower limit of the integration $\nu_0$ is the pion-production threshold on
the nucleon.
A low-energy expansion gives:
\begin{equation}
{\rm Re}[g_{TT}(\nu,Q^2)-g^{pole}_{TT}(\nu,Q^2)]=
(\frac{2\alpha}{M^2})I_{TT}(Q^2)\nu+\gamma_0(Q^2)\nu^3+O(\nu^5).
\end{equation}
Combining Eqs. (1) and (2), the $O(\nu)$ term yields a sum rule 
for the generalized GDH integral\cite{dre01,ji01}:
\begin{eqnarray}
I_{TT}(Q^2) 
&=& \frac {M^2} {4 \pi^2 \alpha} \int_{\nu_0}^{\infty}\frac{K(\nu,Q^2)} {\nu}
\frac{\sigma_{TT}} 
{\nu} d\nu \nonumber \\
 =&& \hspace{-0.5cm} \frac {2M^2} {Q^2} \int_0^{x_0} \Bigr [g_1(x,Q^2) - \frac{4M^2} {Q^2} 
x^2 g_2(x,Q^2)\Bigl ] dx.
\label{eq:gdhsum_def1}
\end{eqnarray}
The low-energy theorem relates I(0) to the anomalous magnetic moment of the nucleon, $\kappa$, and 
Eq.~(\ref{eq:gdhsum_def1}) 
becomes the
original GDH sum rule\cite{GDH}:
\begin{equation}
I(0) =\int_{\nu_0}^{\infty}\frac{\sigma_{1/2}(\nu)-\sigma_{3/2}(\nu)} {\nu}
d\nu
 = -\frac{2 \pi^2 \alpha \kappa^2} {M^2},
\label{eq:gdh}
\end{equation}
where $2 \sigma_{TT} \equiv \sigma_{1/2}-\sigma_{3/2}$.
The $O(\nu^3)$ term yields a sum 
rule for the generalized forward spin polarizability\cite{dre}:
\begin{eqnarray}
\gamma_{TT}(Q^2)&=&
(\frac{1}{2\pi^2})\int^{\infty}_{\nu_0}\frac{K(\nu,Q^2)}{\nu}
\frac{\sigma_{TT}(\nu,Q^2)}{\nu^3}d\nu \nonumber \\
=&&\hspace{-0.5cm}\frac{16 \alpha M^2}{Q^6}\int^{x_0}_0 x^2 \Bigl [g_1(x,Q^2)-\frac{4M^2}{Q^2}
x^2g_2(x,Q^2)\Bigr ] dx. 
\end{eqnarray} 

Considering the longitudinal-transverse interference amplitude $g_{LT}$,
the $O(\nu^2)$ term leads to the generalized longitudinal-transverse
polarizability\cite{dre}:
\begin{eqnarray}
\delta_{LT}(Q^2)&=&
(\frac{1}{2\pi^2})\int^{\infty}_{\nu_0}\frac{K(\nu,Q^2)}{\nu}
\frac{\sigma_{LT}(\nu,Q^2)}{Q \nu^2}d\nu \nonumber \\
=&&\hspace{-0.5cm}\frac{16 \alpha M^2}{Q^6}\int^{x_0}_0 x^2 \Bigl [g_1(x,Q^2)+g_2(x,Q^2)
\Bigr ] dx.   
\end{eqnarray}

Alternatively, we can consider the covariant spin-dependent VVCS amplitudes 
$S_1$ and $S_2$, which are related to the spin-flip amplitudes 
$g_{TT}$ and $g_{LT}$. The unsubtracted
dispersion relations for $S_2$ and $\nu S_2$ lead to a ``super-convergence
relation'' that is valid for any value of $Q^2$,
\begin{equation}
\label{eq:bc}
\int_0^{1}g_2(x,Q^2)dx=0,
\end{equation}
which is the Burkhardt-Cottingham (BC) sum rule\cite{BC}.

At high $Q^2$,
the OPE\cite{JU} for the VVCS amplitude leads to
the twist expansion.
The leading-twist (twist-2) component can be decomposed into flavor triplet ($g_A$), octet ($a_8$) and
singlet ($\Delta\Sigma$) axial charges.
The difference between the proton and the neutron gives the flavor non-singlet
term:
\begin{equation} 
\Gamma_1^p(Q^2)-\Gamma_1^n(Q^2)=\frac{1} {6}g_A+O(\alpha_s)+O(1/Q^2),
\label{eq:genBj}
\end{equation}
which becomes the Bjorken sum rule at the $Q^2\rightarrow \infty$ 
limit. 

The leading-twist part provides information on the polarized parton 
distributions. The higher-twist parts are related to quark-gluon interations
or correlations. 
Of particular interest is 
the twist-3 component, $d_2$, which is related to the second moment of the twist-3 part of 
$g_1$ and $g_2$:
\begin{eqnarray}
 \label{eq:d2}
 d_2(Q^2)
&=& \int_0^1 dx\ x^2 \Bigl (2g_1(x,Q^2)+3g_2(x,Q^2)\Bigr ) \nonumber \\
=&&\hspace {-3mm} 3 \int_0^1 dx\ x^2\Bigl (g_2(x, Q^2)-g_2^{WW}(x,Q^2)\Bigr ),
 \end{eqnarray}
where $g_2^{WW}$ is the twist-2 part of $g_2$ as derived by Wandzura and 
Wilczek\cite{WW}
\begin{equation}
\label{eq:g2ww}
g_2^{WW}(x,Q^2)=g_1(x,Q^2)+\int_x^1 dy \frac{g_1(y,Q^2)} {y}\ .
\end{equation}
$d_2$ is related to the color electric and magnetic polarizabilities, which
describe the response of the collective
color electric and magnetic fields to the spin of the nucleon.
\cite{JU}

\section{Description of the JLab experiments}
\label{JLabexp}

The inclusive experiments described here took place in 
JLab Halls A\cite{HallA nim} and B\cite{HallB nim}. The
accelerator produces a polarized electron beam of energy up to 6 GeV. 

A polarized high-pressure ($\sim$12 atm.) gaseous $^3$He 
target was used as an effective polarized neutron 
target in the experiments performed in Hall A. 
The average target polarization, monitored by NMR and EPR techniques, was 
0.4$\pm 0.014$ and its direction could be oriented longitudinal or transverse 
to the beam direction. 
The measurement of cross sections in the two orthogonal directions allowed 
a direct extraction of $g_1^{3He}$ and $g_2^{3He}$, or 
equivalently $\sigma_{TT}$  and $\sigma_{LT}$.

The scattered electrons were detected by two High Resolution 
Spectrometers (HRS) with the associated detector package. 
The high luminosity of $10^{36}$ cm$^{-2}$s$^{-1}$ allowed for 
statistically accurate data.

The spin structure functions $g_1^n$ and $g_2^n$ are extracted using polarized cross-section differences.
Electromagnetic radiative corrections were performed.
Nuclear corrections are applied via a PWIA-based 
model\cite{cio97}. 
To form the moments, the integrands (e.g. $\sigma_{TT}$ or $g_1$) were 
determined from the measured points by interpolation. 
To complete the moments for the unmeasured high-energy region,
the Bianchi and Thomas parameterization\cite{tho00} was used for 
$4 < W^2 < 1000$ GeV$^2$ and a Regge-type parameterization was used 
for $W^2 > 1000$ GeV$^2$.
Polarized solid $^{15}$NH$_3$ and $^{15}$ND$_3$ targets
using dynamic nuclear polarization were used in Hall B. 
The CEBAF Large Acceptance Spectrometer (CLAS) in Hall B,
which has a large angular 
($2.5 \pi$ sr) and 
momentum acceptance, was used to detect scattered electrons.
The spin structure functions were extracted using asymmetry measurements together with
the world unpolarized structure function fits\cite{F2&RJLab}.
Radiative corrections were applied.

\section{Recent results from Jefferson Lab}
\subsection{Results of the generalized GDH sum and BC sum for $^3$He and the neutron}
\label{GDHsum}
\begin{figure}[!ht]
\parbox[t]{0.45\textwidth}{\centering\includegraphics[bb=80 -28 502 455, angle=0,width=0.45\textwidth]{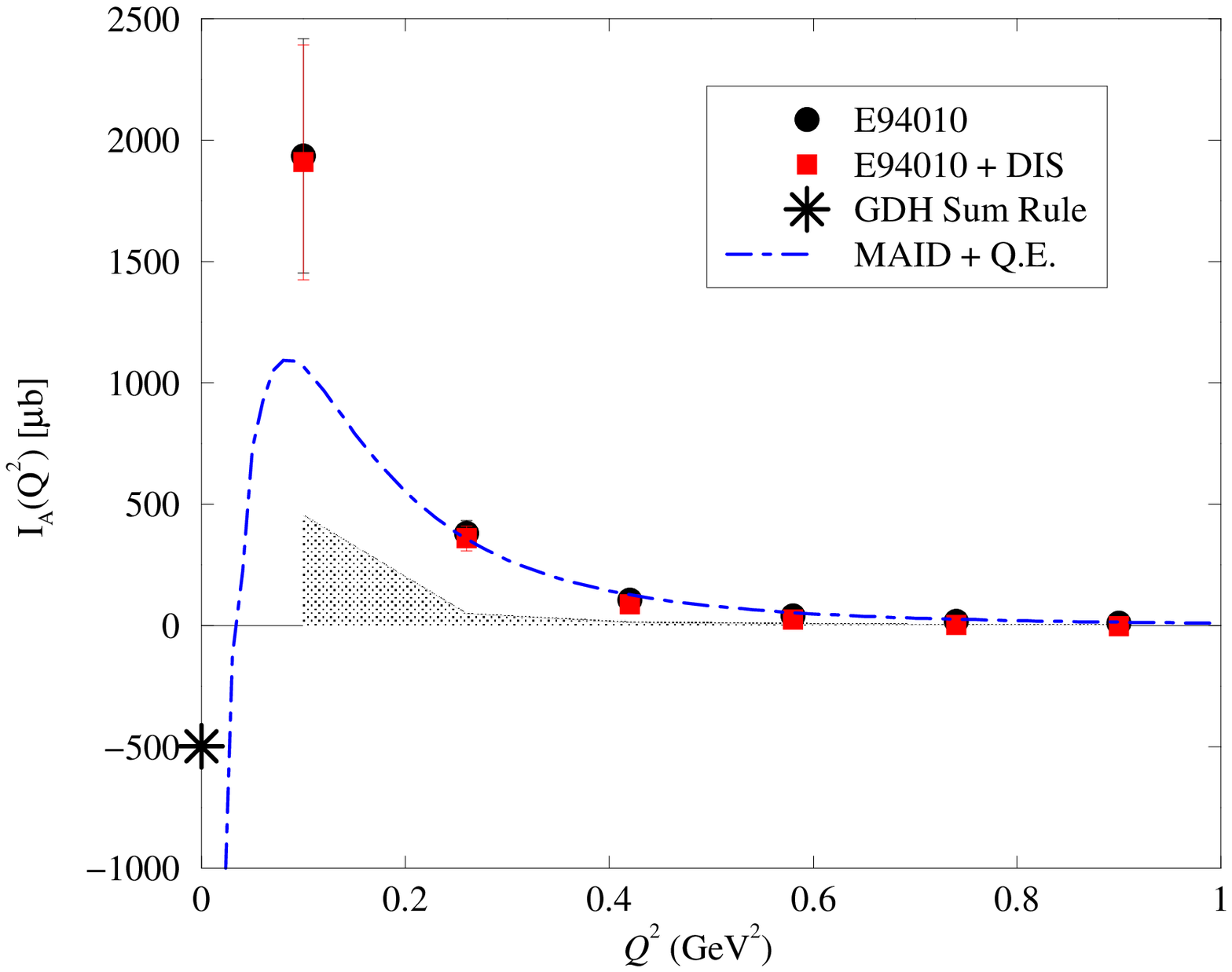}}
\parbox[t]{0.45\textwidth}{\centering\includegraphics[bb=650 100 972 625, angle=-90,width=0.45\textwidth]{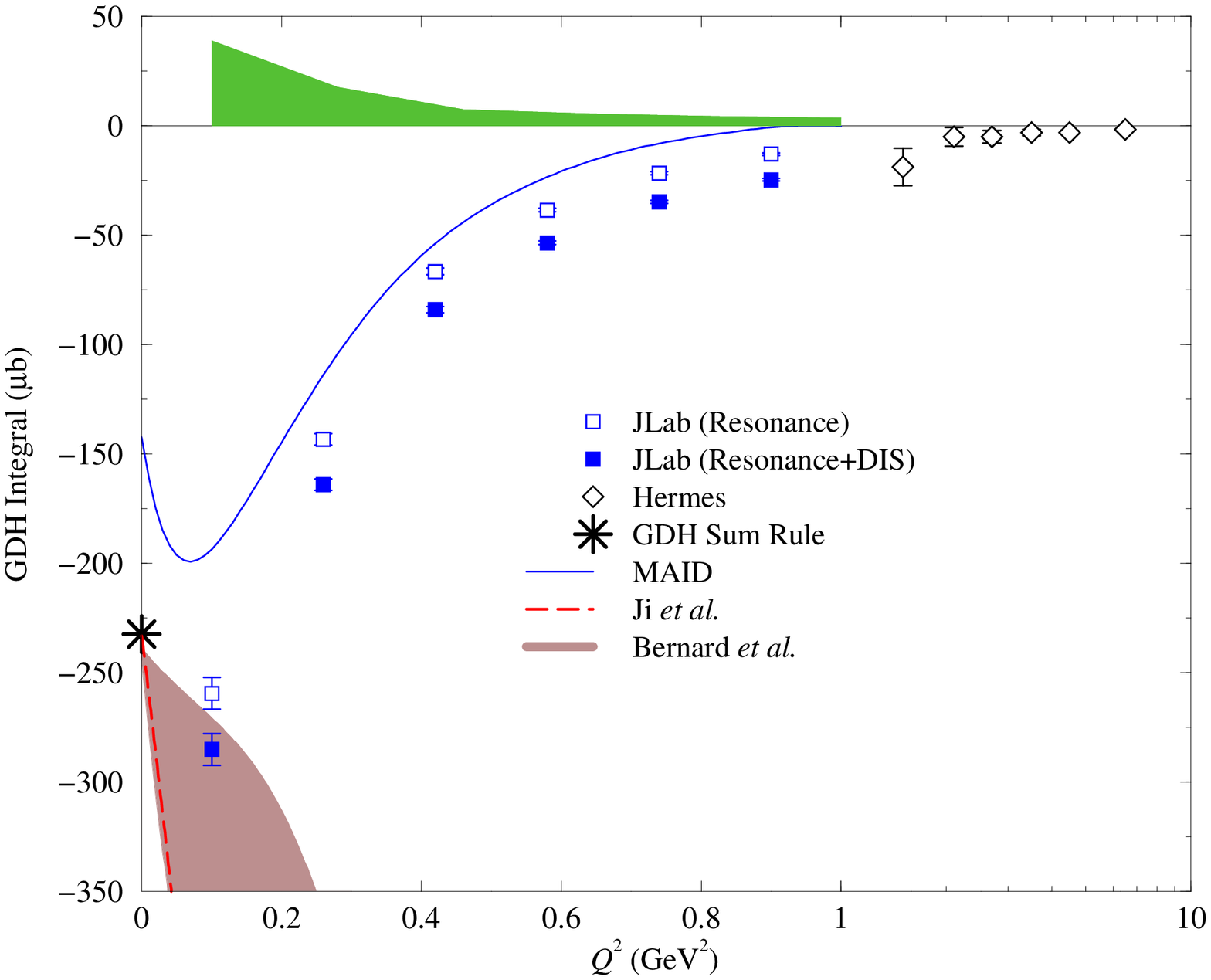}}
\parbox[t]{0.45\textwidth}{\centering\includegraphics[bb=80 -300 502 183, angle=0,width=0.45\textwidth]{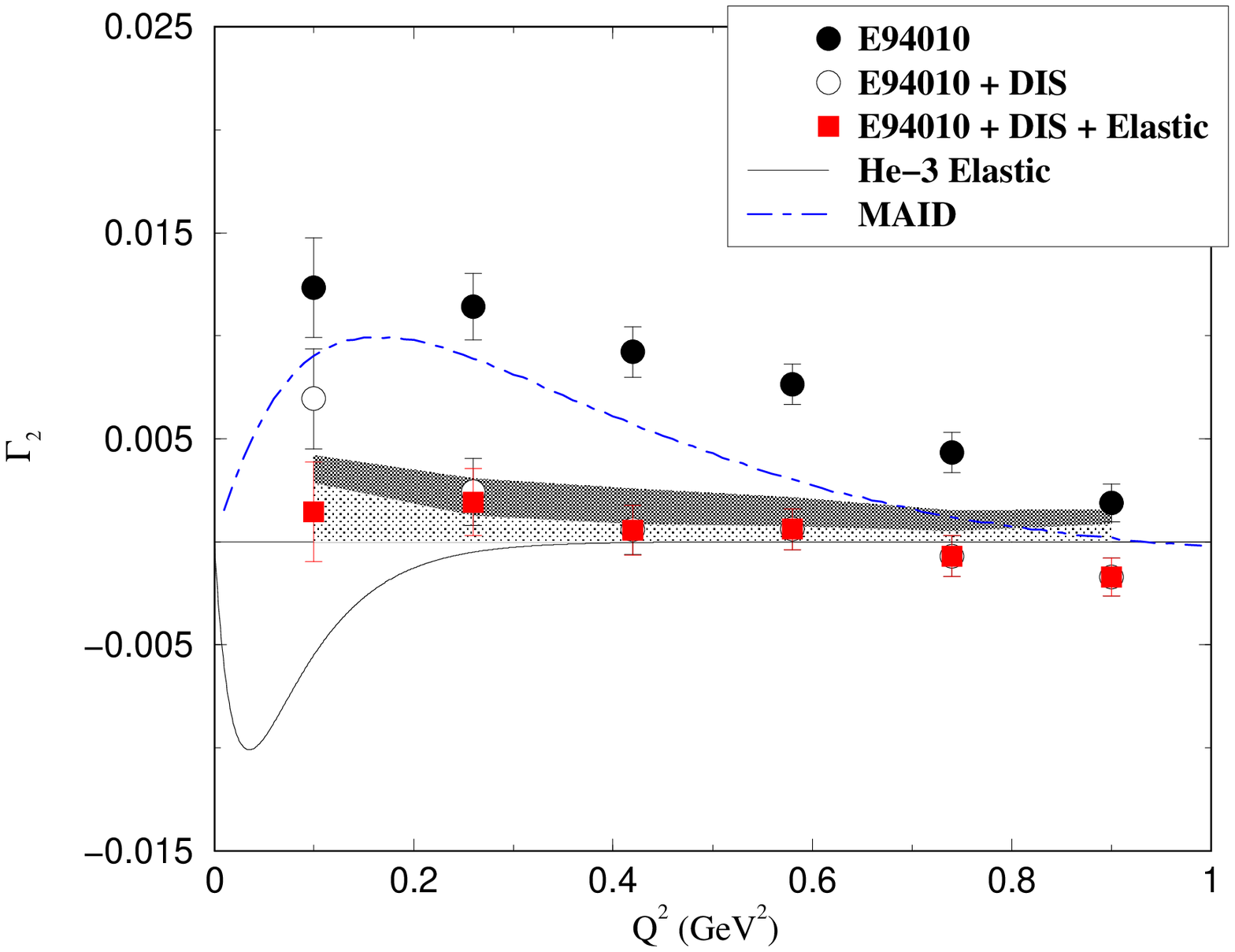}}
\parbox[t]{0.5\textwidth}{\centering\includegraphics[bb=1000 0 1422 600, angle=-90,width=0.5\textwidth]{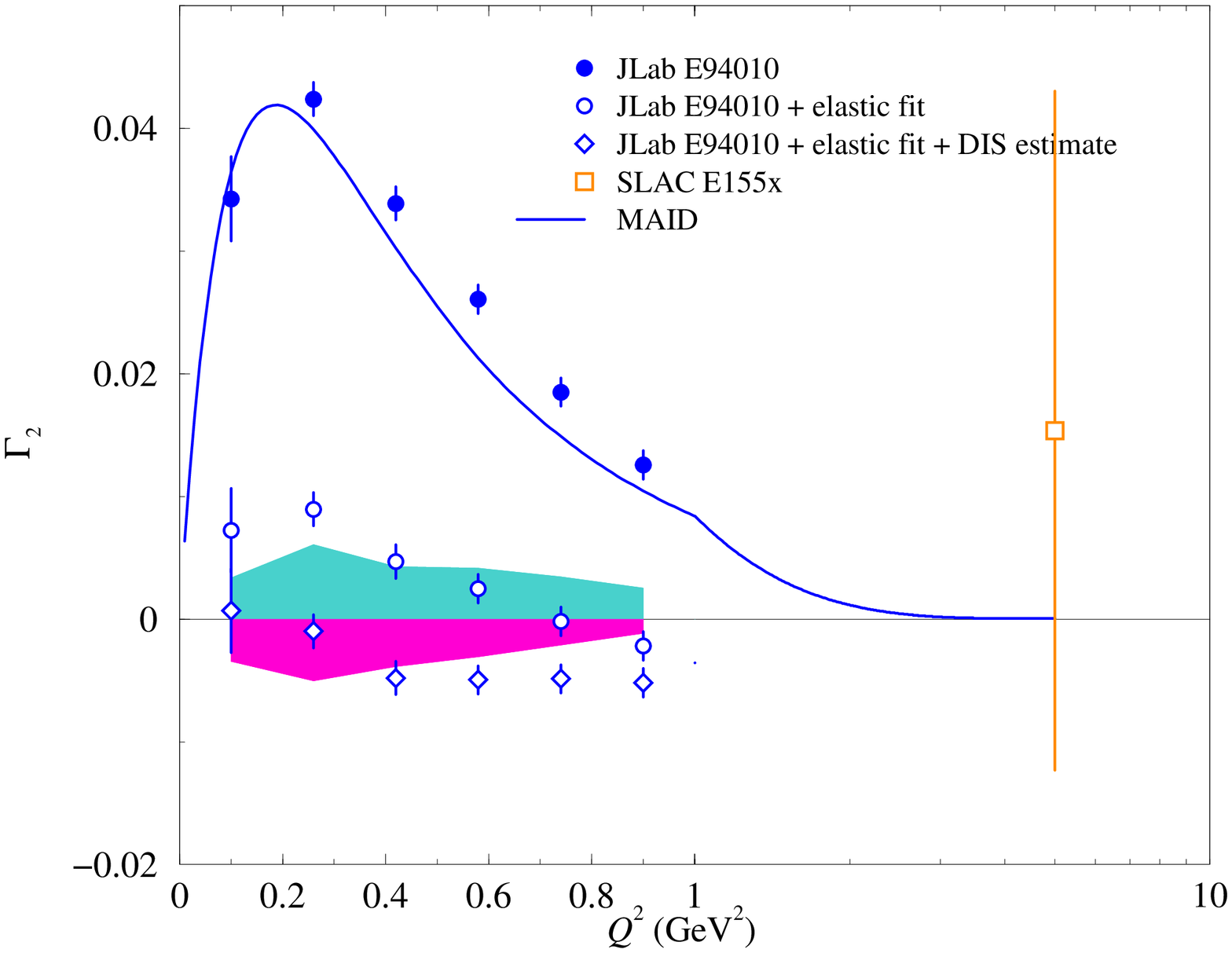}}
\vskip -8cm
\caption{Results of GDH sum $I(Q^2)$ and BC sum $\Gamma_2 (Q^2)$ for $^3$He (preliminary) and the neutron. The $^3$He GDH results are compared with the MAID model plus quasielastic contribution. The neutron GDH results are compared with $\chi$PT calculations of 
ref.~\protect\cite{ji00} (dotted line) and ref.~\protect\cite{ber} 
(dot-dashed line).  
The MAID model calculation of ref.~\protect\cite{dre01},
is represented by a solid line.  Data from HERMES\protect\cite{HERMES} are also shown. The BC sum results (resonance only) are compared with MAID model calculations.}
\label{fig:GDH}
\end{figure}

Fig.~\ref{fig:GDH} shows the extended GDH integrals  
$I(Q^2)$ (open circles) for $^3$He (preliminary) (upper-left) and  for the neutron (upper-right),
which were extracted from Hall A experiment E94-010\cite{e94010} , from break-up threshold for $^3$He (from pion threshold for the neutron) to $W=2$ GeV.
The uncertainties, when visible, represent statistics only; the systematics are shown by the grey band.   
The solid squares include an estimate of the unmeasured high-energy part.
The corresponding
uncertainty is included in the systematic uncertainty band.
The preliminary $^3$He results rise with decreasing $Q^2$. Since the GDH sum 
rule at $Q^2=0$ predicts a large negative value, a drastic turn around 
should happen at $Q^2$ lower than 0.1 GeV$^2$. A simple model using MAID\cite{dre01} plus quasielastic contributions indeed shows the expected turn
around. The data at low $Q^2$ should be a good test ground for few-body Chiral Pertubation Theory Calculations.

The neutron results indicate a smooth variation of $I(Q^2)$  to increasingly negative values as $Q^2$ varies from $0.9\,{\rm GeV^2}$ 
towards zero.
The data are more negative than the MAID model calculation\cite{dre01}. 
Since the calculation only includes contributions to $I(Q^2)$ for $W \le 2\,{\rm GeV}$, it should be compared with the
open circles. The GDH sum rule 
prediction, $I(0)=-232.8\,\mu{\rm b}$, is indicated in Fig.~\ref{fig:GDH}, along with  
extensions to $Q^2>0$ using two next-to-leading order $\chi$PT
calculations, one using the Heavy Baryon approximation (HB$\chi$PT) \cite{ji00} 
(dotted line) and the other Relativistic Baryon $\chi$PT (RB$\chi$PT)\cite{ber} (dot-dashed line). Shown with a grey band is RB$\chi$PT including resonance effects\cite{ber}, which have an associated
large uncertainty due to the resonance parameters used. 

 The capability of transverse polarization of the Hall A $^3$He target 
allows precise measuremetns of $g_2$. 
The integral of $\Gamma_2^{^3He}$ (preliminary) and $\Gamma_2^n$ 
is plotted in the lower-left and lower-right panels of 
Fig.~\ref{fig:GDH} in the measured region (solid circles) and open circles show
the results after adding an estimated
DIS contribution for $^3$He (elastic contribution for the neutron). 
The solid squares (open diamonds)
correspond to the results obtained after adding the elastic contributions for $^3$He,
(adding an estimated DIS 
contribution assuming $g_2 = g_2^{WW}$ for the neutron). The MAID estimate agrees with the general trend but slightly lower than the resonance data. 
The two bands correspond to the experimental systematic
errors and the estimate of the systematic error 
for the low-$x$ extrapolation. The total results are consistent with the BC 
sum rule. 
The SLAC E155x collaboration\cite{SLAC} previously reported 
a neutron result at high $Q^2$ (open square), which is consistent with zero but with a rather 
large error bar. On the other hand, the SLAC proton result was reported to deviate 
from the BC sum rule by 3 standard deviations. 
\subsection{First moments of $g_1$ and the Bjorken sum}
\begin{figure}[ht!]
\parbox[t]{0.43\textwidth}{\centering\includegraphics[bb=80 17 502 500, angle=0,width=0.43\textwidth]{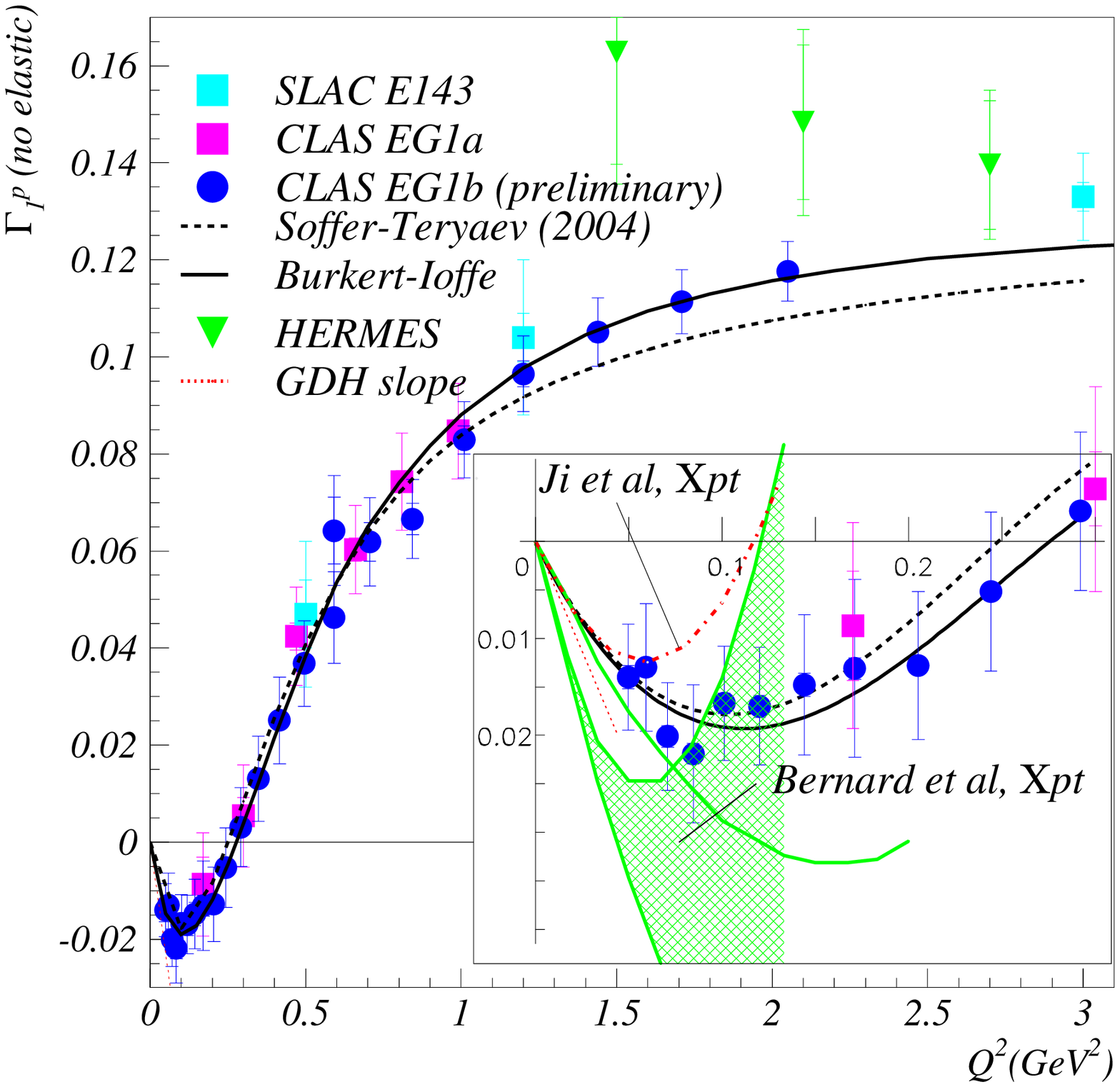}}
\parbox[t]{0.43\textwidth}{\centering\includegraphics[bb=80 17 502 500, angle=0,width=0.43\textwidth]{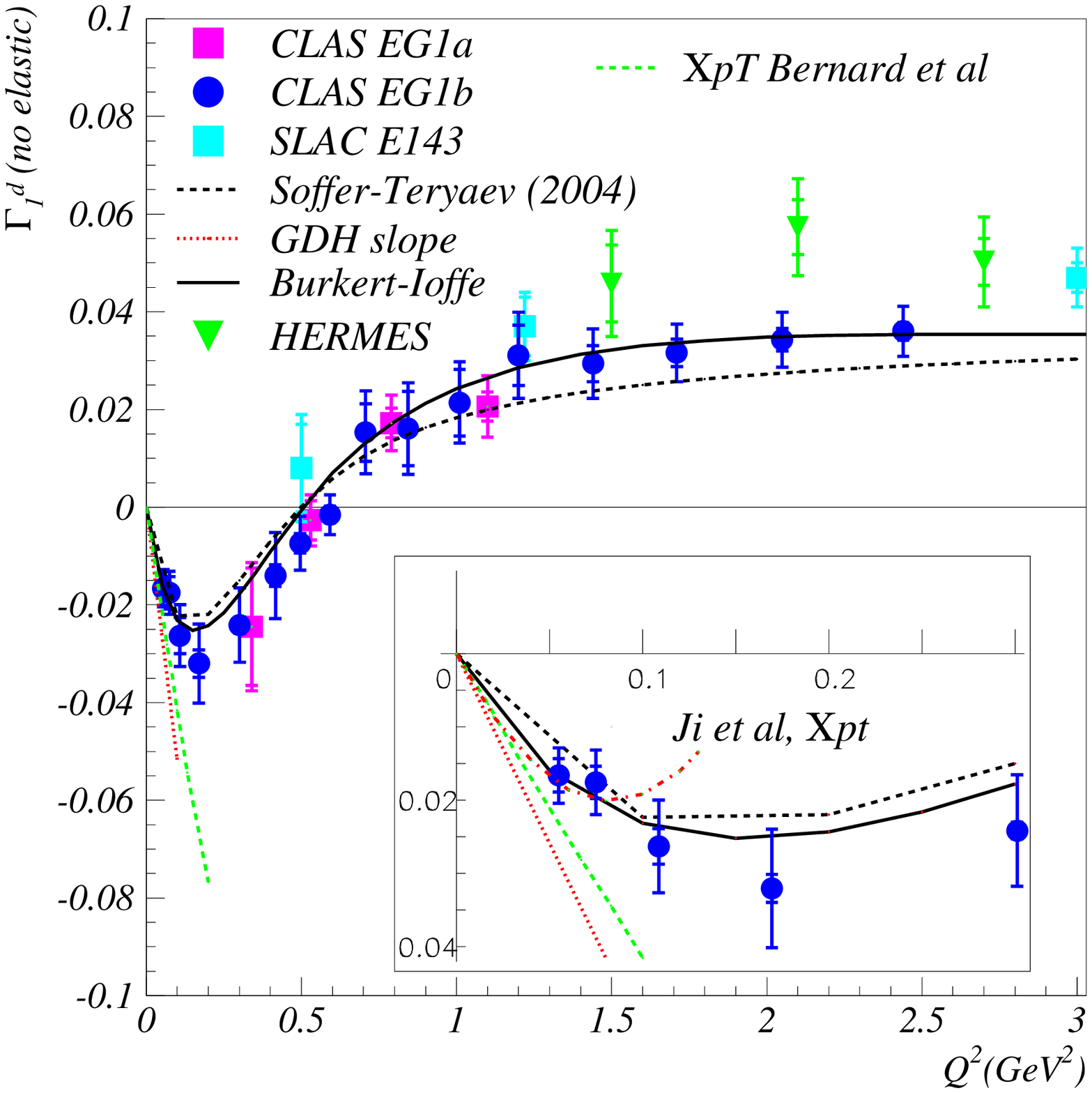}}
\parbox[t]{0.43\textwidth}{\centering\includegraphics[bb=80 37 502 520, angle=0,width=0.43\textwidth]{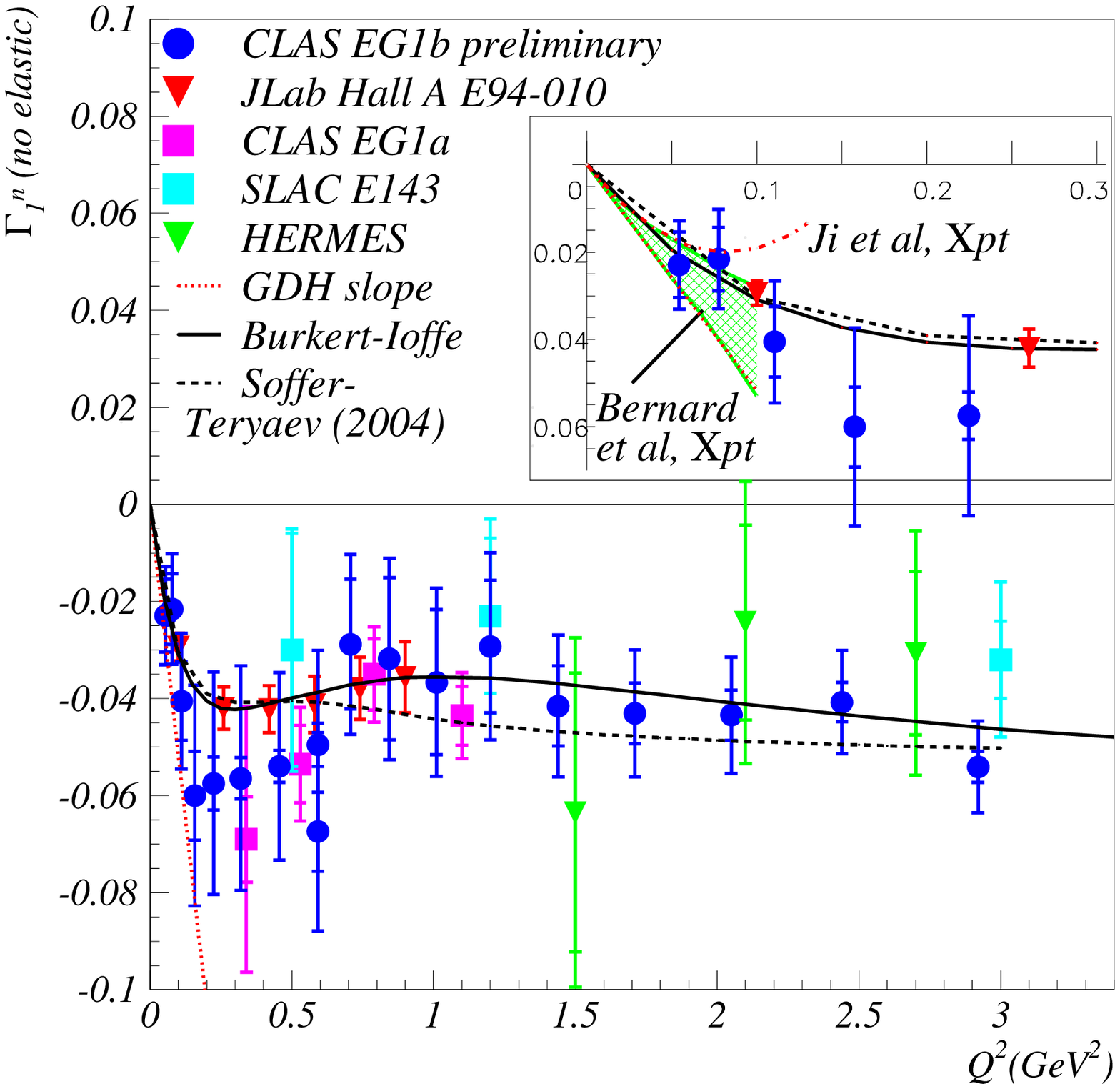}}
\parbox[t]{0.42\textwidth}{\centering\includegraphics[bb=-35 50 367 533, angle=0,width=0.42\textwidth]{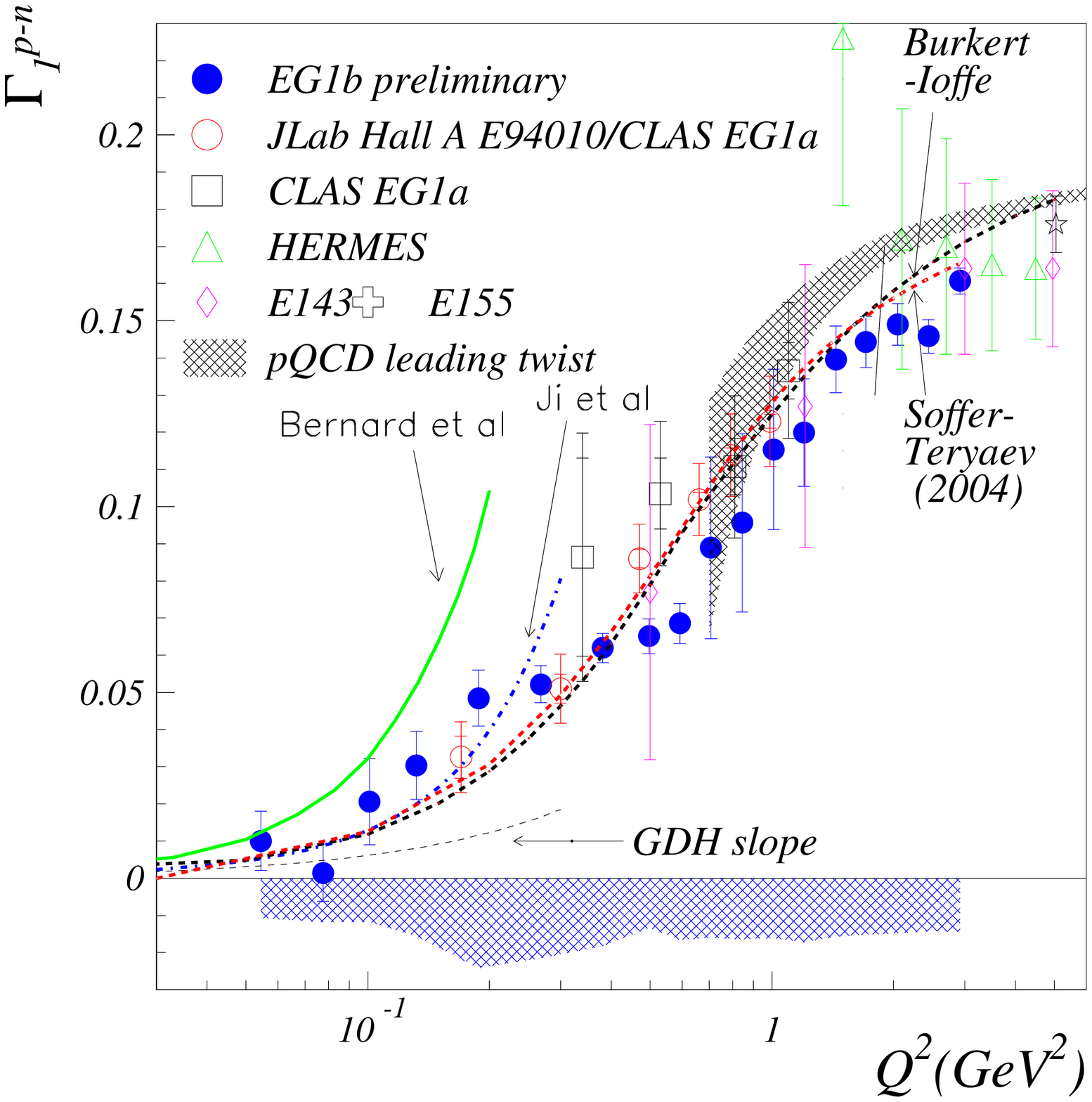}}
\vspace {3mm}
\caption{Preliminary results of $\Gamma_1 (Q^2)$ from CLAS eg1b\protect\cite{eg1b} for p, d, n and p-n, together with the published results from Hall A\protect\cite{e94010} and CLAS eg1a\protect\cite{eg1p,eg1d} .
The slopes at $Q^2$=0 predicted by the GDH sum rule 
are given by the dotted lines. The MAID model  
predictions that include only resonance contributions are shown by the full 
lines while 
the dashed (dot-dashed) lines are the predictions from the Soffer-Teryaev (Burkert-Ioffe) model.
The leading twist $Q^2$-evolution of the moments is given by the grey band. 
 The insets show comparisons with $\chi$PT calculatiosns. The full
lines (bands) at low $Q^2$ are the next-to-leading order $\chi$PT 
predictions by Ji \emph{et al.} (Bernard \emph{et al.}). } 
\label{fig:gamma1pn}
\vspace {-4mm}
\end{figure}

The preliminary results from Hall B EG1b\cite{eg1b} experiment on $\bar \Gamma_1(Q^2)$ at low to moderate $Q^2$ are shown together with published results from Hall A\cite{e94010} and Hall B eg1a\cite{eg1p,eg1d} in Fig.~\ref{fig:gamma1pn} along with the data from 
SLAC\cite{SLAC} and HERMES\cite{HERMES}.
The new results are in good agreement with the published data.
 The
inner uncertainty
indicates the statistical uncertainty while the outer one is the quadratic sum of the 
statistical and systematic uncertainties. 

At $Q^2$=0, the GDH sum rule predicts the slopes of moments (dotted lines). The deviation from the
slopes at low $Q^2$ can be calculated with $\chi$PT. We show results of calculations by Ji {\it et al.}\cite{ji00} 
using HB$\chi$PT and by Bernard {\it et al.} with and without\cite{ber} the 
inclusion of vector 
mesons and $\Delta$ degrees of freedom. 
The calculations are in reasonable agreements with the data at the lowest $Q^2$
settings of 0.05 - 0.1 GeV$^2$.
At moderate and large $Q^2$ data are compared with two model calculations~\cite{sof02,bur92} . Both models agree well with the data.  
The leading-twist pQCD evolution is shown by the grey band. It tracks the data down to
surprisingly low $Q^2$, which indicates an overall suppression of higher-twist
effects. 

\subsection{Spin Polarizabilities: $\gamma_0$, $\delta_{LT}$ and $d_2$ for the neutron}
The generalized spin polarizabilities provide benchmark tests
of $\chi$PT calculations at low $Q^2$.
Since the generalized polarizabilities have an extra $1/\nu^2$ 
weighting compared to the first moments (GDH sum or 
$I_{LT}$), these integrals have less contributions from the large-$\nu$ 
region and converge much faster, which minimizes the uncertainty due to
the unmeasured region at large $\nu$. 

At low $Q^2$, the 
generalized polarizabilities have been evaluated with next-to-leading order 
$\chi$PT 
calculations\cite{van02,ber}.
One issue in the $\chi$PT calculations is how to properly
include the nucleon resonance contributions, especially the $\Delta$ resonance.
As was pointed out in Refs.~\cite{van02,ber} , while $\gamma_0$ is sensitive to 
resonances, $\delta_{LT}$ is insensitive to the $\Delta$ 
resonance. Measurements of the generalized spin
polarizabilities are an important step in understanding the dynamics of
QCD in the chiral perturbation region.

The first results for the neutron generalized forward
spin polarizabilities $\gamma_0(Q^2)$ and $\delta_{LT}(Q^2)$
were obtained at Jefferson Lab Hall A\cite{e94010}.
\begin{figure}[!ht]
\parbox[t]{0.43\textwidth}{\centering\includegraphics[bb=80 130 502 630, angle=0,width=0.43\textwidth]{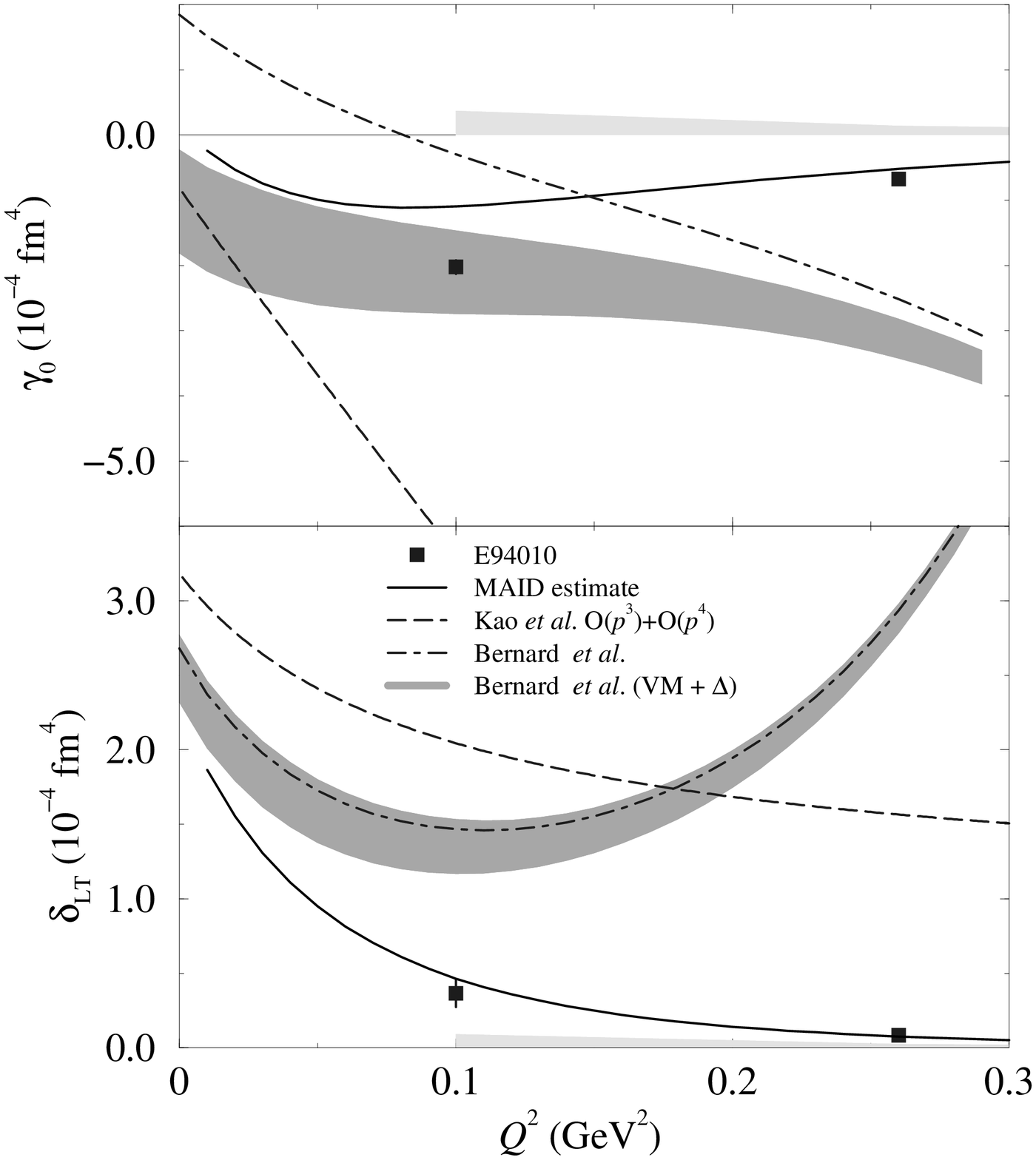}}
\parbox[t]{0.45\textwidth}{\centering\includegraphics[bb=-74 17 372 500, angle=0,width=0.45\textwidth]{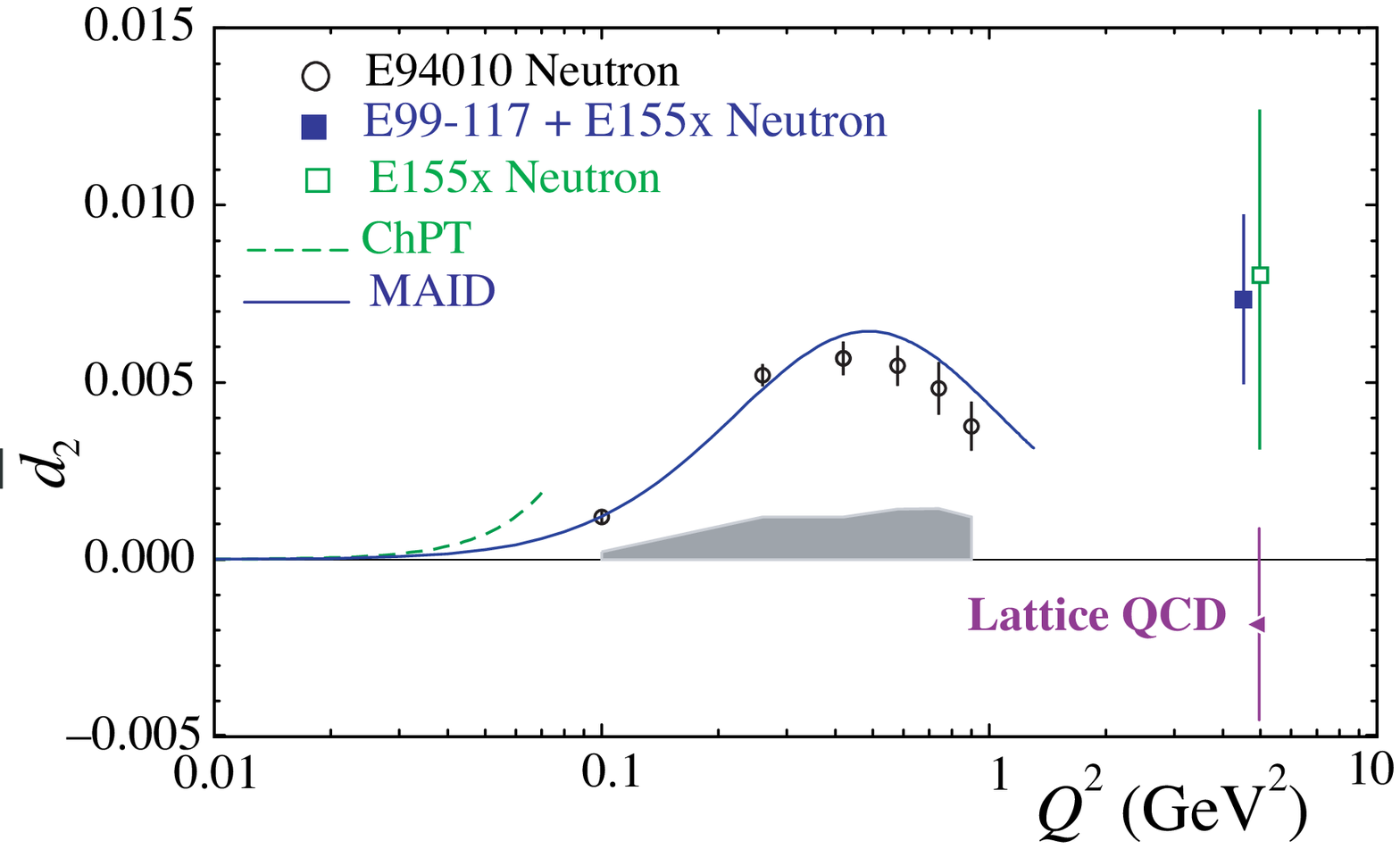}}

\vspace {10mm}
\caption{Results for the neutron spin polarizabilities 
$\gamma _0$ (top-left panel) and $\delta _{LT}$
(bottom-left panel). Solid squares represent the results with statistical uncertainties.
The light bands represent the systematic uncertainties. 
The dashed curves represent the HB$\chi $PT 
calculation\protect\cite{van02}. 
The dot-dashed curves and the dark bands represent the 
RB$\chi $PT calculation 
without and with\protect\cite{ber} 
the $\Delta $ and vector meson contributions, respectively.
Solid curves represent the MAID model\protect\cite{dre01}.
The right panel shows the $\bar d_2^n$ results from JLab\protect\cite{e94010,e99117} and 
SLAC~\protect\cite{SLAC}, together with 
the Lattice QCD calculations~\protect\cite{goc01}. 
}
\label{fig:fig3}
\vspace {-3mm}  
\end{figure}
The results for $\gamma_0(Q^2)$ are shown
in the top-left panel of Fig.~\ref{fig:fig3}. 
The statistical uncertainties are smaller than the
size of the symbols. 
The data are compared with 
a next-to-leading order ($O(p^4)$) HB$\chi$PT 
calculation\cite{van02}, a next-to-leading order RB$\chi$PT
calculation and the same calculation explicitly including 
both the $\Delta$ resonance and vector meson contributions\cite{ber}.
Predictions from the MAID model\cite{dre01} are also shown.
At the lowest $Q^2$ point,
the RB$\chi$PT
calculation including the resonance contributions
is in good agreement with the experimental result.
For the HB$\chi$PT calculation without explicit resonance contributions, 
discrepancies are large even at $Q^2 = 0.1$ GeV$^2$. 
This might indicate the significance of the resonance contributions or a
problem with the heavy baryon approximation at this $Q^2$.
The higher $Q^2$ data point is in good agreement with the MAID
prediction,
but the lowest data point at $Q^2 = 0.1 $ GeV$^2$ is significantly lower.
Since $\delta_{LT}$ is 
insensitive to the
$\Delta$ resonance contribution, it was believed that $\delta_{LT}$ should be
more suitable than $\gamma_0$ to serve as a testing ground for the chiral 
dynamics of QCD~\cite{ber,van02}.
The bottom-left panel of Fig.~\ref{fig:fig3} shows $\delta_{LT}$ 
compared to
$\chi$PT calculations and the MAID predictions. While the MAID predictions are in good agreement with the results, it is surprising to see
that 
the data are in significant disagreement with the $\chi$PT calculations 
even at the lowest $Q^2$, 0.1 GeV$^2$. 
This disagreement presents a significant challenge to the present Chiral Pertubation Theory.

New experimental data have been taken at very low $Q^2$, down to 0.02 GeV$^2$
for the neutron ($^3$He)\cite{e97110} and the proton and deuteron\cite{eg4}.
Analyses are underway. Preliminary asymmetry results just became available for the neutron. These results will shed light and provide benchmark tests to the
$\chi$PT calculations at the kinematics where they are expected to work.

Another combination of the second moments, $d_2(Q^2)$, provides an 
efficient way to study the high $Q^2$ behavior of the nucleon spin structure,
since it is a matrix element, related to the color polarizabilities and can 
be calculated from Lattice QCD. It also provides a means to study the 
transition from high to low $Q^2$. 
In Fig.~\ref{fig:fig3}, $\bar d_2(Q^2)$  
is shown. The experimental results are the solid circles. 
The grey band represents the systematic uncertainty. The world neutron results from SLAC\cite{SLAC}(open square) and from JLab E99-117\cite{e99117} 
(solid square) are also shown. The solid line is the
MAID calculation containing only the resonance contribution. 
At low $Q^2$ the HB$\chi$PT calculation\cite{van02} (dashed line) is shown.
The RB$\chi$PT
with or without the vector mesons  
and the $\Delta$ contributions\cite{ber} 
are very close to the HB$\chi$PT curve at this scale, and are not shown on the
figure for clarity.
The Lattice QCD prediction\cite{goc01} at $Q^2$ = 5~GeV$^2$ is negative but close to zero. There is a $2\sigma$ deviation from the experimental result.
We note that all models (not shown at
this scale) predict a negative or zero value at large $Q^2$. 
At moderate $Q^2$,
our data show that $\bar d_2^n$ is positive and decreases with $Q^2$.

Preliminary results at a $Q^2$ range of 1-4 GeV$^2$ for the neutron\cite{e01012} are available now. 
New experiments are planned with 6 GeV beam\cite{d26gev} at average 
$Q^2$ of 3~GeV$^2$ and with 
future 12 GeV upgraded JLab\cite{d212gev} at constant $Q^2$ values of 3, 4 and 
5~GeV$^2$. They will provide a benchmark test of the lattice QCD
calculations.

\section{Conclusion}

A large body of nucleon spin-dependent cross-section and 
asymmetry data have been collected at low to moderate $Q^2$ in the resonance region. 
These data have been used to evaluate the $Q^2$ evolution of moments of 
the nucleon spin structure functions $g_1$ and $g_2$, including the 
GDH integral, the Bjorken sum, the BC sum and the spin polarizabilities. 
  
At $Q^2$ close to zero, available next-to-leading order $\chi$PT calculations
were tested against the data and found to be in reasonable agreement for $Q^2$ 
of 0.05 to
0.1 GeV$^2$ for the GDH integral $I(Q^2)$, $\Gamma_1(Q^2)$ and the  forward spin
polarizability $ \gamma_0(Q^2)$. Above $Q^2$ of 0.1 GeV$^2$  a significant 
difference between the calculation and the data is observed, pointing to the
limit of applicability of $\chi$PT as $Q^2$ becomes larger. Although it was
expected that the $\chi$PT calculation of $\delta_{LT}$ would offer a faster
convergence because of the absence of the $\Delta$ contribution, the
experimental data show otherwise. None of the available $\chi$PT
calculations can  reproduce $\delta_{LT}$ at $Q^2$ of 0.1 GeV$^2$.  This discrepancy presents a significant challenge
to our theoretical understanding at its present level of approximations. 

Overall, the trend of the data  is well described by phenomenological 
models. The dramatic $Q^2$ evolution of $I_{GDH}$ from high to  low $Q^2$ was
observed as predicted by these models for both the proton and the neutron. This 
behavior is mainly determined by the relative strength and sign of the $\Delta$ 
resonance compared to that of higher-energy resonances and deep inelastic processes.
This also shows that the current level of phenomenological understanding of the
resonance spin structure using these moments as observables is reasonable. 

The BC sum rule for both the neutron and $^3$He is observed to be satisfied 
within
uncertainties due to a cancellation between the resonance and the
elastic contributions. The BC sum rule is expected to be valid at all $Q^2$.
This test validates the assumptions going into the BC sum rule,
which provides confidence in sum rules with similar assumptions.

Overall, 
the recent JLab data have provided valuable information on the transition between the non-perturbative 
to the perturbative regime of QCD. They form a precise data set for a check of $\chi$PT calculations. 
New results at very low $Q^2$ for the neutron\cite{e97110}, proton and 
deuteron\cite{eg4} will be 
available soon. They will provide benchmark tests of the Chiral Pertubation
Theory calculations in the kinematical region where they are expected to
work. 

Future precision measurements\cite{d26gev,d212gev} of $d_2^n$
 at $Q^2 = 3-5$~GeV$^2$
will provide a benchmark test of Lattice QCD.

\section*{Acknowledgments}
This work was supported by the U.S. Department of Energy (DOE).
The Southeastern Universities
Research Association operates the Thomas Jefferson National Accelerator
Facility for the DOE under contract DE-AC05-84ER40150, Modification No. 175.





\end{document}